\documentclass[aps,pra,twocolumn,amsmath,amssymb,nofootinbib,showpacs,superscriptaddress]{revtex4-1}
\usepackage[english]{babel}
\usepackage{latexsym}
\usepackage{graphics}
\usepackage{graphicx}
\usepackage{epsfig}
\usepackage{color}
\usepackage{bm}
\usepackage{amsmath}
\usepackage{amssymb}
\usepackage{amsthm}
\usepackage{dcolumn}
\usepackage{bm}
\usepackage{float}
\usepackage{hyperref}
\usepackage{color}
\usepackage{epstopdf}
\usepackage{braket}
\usepackage{cleveref}
\usepackage[svgnames]{xcolor}
\hypersetup{hidelinks,colorlinks=true,allcolors=DarkBlue}

\theoremstyle{statement}
\newtheorem{statement}{Statement}

\theoremstyle{conjecture}
\newtheorem{conjecture}{Conjecture}

\begin{document}

\preprint{APS/123-QED}

\title{What is computable and non-computable in the quantum domain: \\ Statements and conjectures}

\author{Aleksey K. Fedorov}
\affiliation{P.N. Lebedev Physical Institute of the Russian Academy of Sciences, Moscow 119991, Russia}
\affiliation{Russian Quantum Center, Skolkovo, Moscow 121205, Russia}
\affiliation{National University of Science and Technology ``MISIS'', Moscow 119049, Russia}

\author{Evgeniy O. Kiktenko}
\affiliation{Russian Quantum Center, Skolkovo, Moscow 121205, Russia}
\affiliation{National University of Science and Technology ``MISIS'', Moscow 119049, Russia}

\author{Nikolay N. Kolachevsky}
\affiliation{P.N. Lebedev Physical Institute of the Russian Academy of Sciences, Moscow 119991, Russia}
\affiliation{Russian Quantum Center, Skolkovo, Moscow 121205, Russia}

\date{\today}
\begin{abstract}
Significant advances in the development of computing devices based on quantum effects and the demonstration of their use to solve various problems have rekindled interest in the nature of the ``quantum computational advantage''. 
Although various attempts to quantify and characterize the nature of the quantum computational advantage have previously been made, this question largely remains open. 
Indeed, there is no universal approach that allows determining the scope of problems whose solution can be accelerated by quantum computers, in theory of in practice. 
In this paper, we consider an approach to this question based on the concept of complexity and reachability of quantum states. 
On the one hand, the class of quantum states that are of interest for quantum computing must be {\it complex}, i.e., not amenable to simulation by classical computers with less than exponential resources. 
On the other hand, such quantum states must be {\it reachable} on a practically feasible quantum computer. 
This means that the unitary operation that transforms the initial quantum state into the desired one must be decomposable into a sequence of one- and two-qubit gates of a length that is at most polynomial in the number of qubits. 
By formulating several statements and conjectures, we discuss the question of describing a class of problems whose solution can be accelerated by a quantum computer.
\end{abstract}

\maketitle

\section{Introduction}
Developments in technology of producing processors based on semiconductor platform~\cite{Moore1965}  have made it possible to constantly increase the computational power during last decades and have brought the society to the era, in which computational devices are used on the daily basis.
However, certain computational problems remain of exceptionally high computational complexity for such computing devices. 
Examples include integers factorization, in which one needs for a given general composite number $N$ find its nontrivial factors $p$ and $q$ (such that $N=p{\times}q$), simulating complex quantum systems, in particular, calculating energy states of large-scale molecules, 
and combinatorial optimization problems, where one has to find the best solution among a large set of possible candidates.
Although classical\footnote{We say that something is ``classical'' when one can describe its properties without invoking quantum phenomena like entanglement} computational devices continue their developments, and the ``death of the Moore's law''~\cite{Waldrop2016} is rather questionable~\cite{Fedorov2022}, aforementioned computational problems cannot be efficiently solved via existing devices and known algorithmic approaches.

One of the ways to extend computational capabilities it to build a drastically different type of quantum devices, which use phenomena that take place at the level of individual quantum objects, like single atoms or photons~\cite{Brassard1998}.
Such quantum computational devices are considered as useful in solving various classes of computational problems, which are known to be classically hard; this list includes aforementioned integer factorization~\cite{Shor1994} and simulating quantum systems~\cite{Lloyd1996}, which are believed to be solvable with quantum computers in polynomial time.  

Historically, the Feynman's approach to building quantum computers is to use such devices to simulate other quantum systems~\cite{Feynman1982,Feynman1986} that are believed to be hard to simulate using classical resources. 
Specifically, it is conjectured that for a class of quantum many-body systems resources to simulate it, i.e., predict the probability of a given measurement outcome (a so-called strong simulation) or mimic a sampling measurement outcomes (weak simulation), grows exponentially with the system size.
This is known as the {\it entanglement frontier}~\cite{Preskill2012}.
For example, let us consider a state of $n$ qubits, where each of the qubit is described by the state $\ket{\psi}=\bigotimes_{i}(\alpha_i\ket{0}+\beta_i\ket{1})$, 
where $\alpha_i$ and $\beta_i$ are complex numbers with the normalization condition $|\alpha_i|^2+|\beta_i|^2=1$ for $i=1,\ldots,n$.
The description of such a state requires up to $2n$ real numbers, if, e.g., Bloch sphere angles are used for parametrization.
However, if we apply a long enough sequence of single-qubit operations  and two-qubit entangling  operations, e.g., the ${\rm CNOT}$ gate~\cite{Ladd2010}, between different pairs of qubits, 
then its many-body quantum state becomes entangled. 
It is then seems that there will no obvious way to simulate such a state with linear resources. 
One may conjecture that we would need resources scaling as $2^n$, i.e. exponentially with the system size. 
For the $n=100$ case, a straightforward simulation requires storing $2^{100}$-dimensional complex vector in the memory and calculating the results of rotations in $2^{100}$ size space, which seems to be impossible with the use of any foreseen classical computing device.
From this perspective, having any entangled superposition state can be considered as the condition for achieving quantum computational advantage. 

However, the celebrated Gottesman-Knill theorem~\cite{Gottesman1998,NielsenChuang2000,Gottesman2004} says that sometimes it is not the case. 
A many-body entangled state that is prepared by set of gates belonging to the Clifford set applied to a computational basis state, the so-called stabilizer state, is amenable to polynomial-resource strong simulation with respect to any Pauli measurement, including the computational basis measurement. 
An example of entangled many-body quantum states that belongs to this class are the Greenberger-Horne-Zeilinger (GHZ) state and graph states.
This consideration shows that a way of thinking about quantum computers as devices, whose computational advantage is related \emph{only} to the entangled superposition nature of the underlying quantum state, is naive. 
Another example is quantum circuits consisting matchgates, which is known to be efficiently simulatable on a classical computer~\cite{Ermakov2024}.

For achieving quantum computational advantage indeed the question of the simulatability of quantum states plays an important role, but it is related not only to the amount of quantum entanglement, but also to a type of readout measurement and employed gates.
Indeed, applying a layer of non-Clifford gates to an entangled stabilizer state just before the computational basis measurement, which is equivalent to making some general local measurement, does not change state's entanglement, but removes applicability of the Gottesman-Knill theorem.
Similarly, adding non-Clifford gates, such as T-gate, in a quantum circuit makes it non-simulable within classical resources.
These simple examples show that the space of possible states of quantum systems --- its Hilbert space --- is not uniform from the viewpoint of complexity of their simulation (with respect to computational basis measurement): $n$-qubit separable states require linear resources, 
$n$-qubit entangled states that are prepared only via Clifford operations are simulatable with polynomial complexity, whereas we also know $n$-qubit entangled quantum states, e.g. prepared with non-Clifford operations, that are required exponential resources to be simulated. 

On the other hand, although quantum processor can be considered as a universal computational device, i.e. in principle one can realize arbitrary unitary operation (up to a prespecified  precision), not all the unitary operations can be efficiently decomposed on the sequence of single- and two-qubit quantum operations that are practically realizable on quantum processors. 
Indeed, if one considers an arbitrary $2^{n}\times2^{n}$ unitary, typically for its decomposition one needs a sequence of 2$\times$2 and $4\times{4}$ unitaries that is exponentially long. 
For several exceptional cases the length of such sequence can be linear or polynomial as in the case of Shor's factorization algorithm~\cite{Shor1994}.
From this Manin's perspective~\cite{Manin1980}, not all quantum states are {\it reachable} with a practical quantum computer.

Thus, quantum computing devices are helpful for classically non-simulatable and quantum-reachable states.
How large is such set of quantum states?
How such set of quantum states is structured in a Hilbert space?
These questions motivate us to make first attempts to classify quantum states from the aforementioned points of view.
Finding relations between different types of quantum states may shine a new light on the origins of quantum computational advantage, which in this context follows from the size and structure of the set of ``complex'' quantum states that are not simulatable classically with less that exponential resources.
We present a sketch of the quantum state complexity diagrams, where exact relations between certain classes of quantum states are not yet exactly defined.
Our classification takes into account existing ways to quantify complexity of quantum states coming from quantum many-body physics, condensed matter, and quantum information theory.

\section{Quantum state picture and universal gate-based quantum computing model}

The main object of interest in our case is an $n$-qubit quantum state $\ket{\Psi_n}$, which appears after execution of all the unitaries applied to $\ket{0}^{\otimes n}$ in a quantum circuit, i.e. the quantum state appearing before computational basis measurements. 
From a different prospective, one may think not about complex quantum states, but complex quantum processes, complex Hamiltonians that govern non-trivial quantum dynamics, or complex measurements. 
In the first case, the Jamiolkowski Isomorphism allows considering all quantum processes, i.e. ones correspond to quantum circuits, as quantum states in the extended Hilbert space.
In the case of complex Hamiltonians and measurements, the equivalence between adiabatic~\cite{Aharonov2007}, measurement-based~\cite{Briegel2001}, and gate-based~\cite{Ladd2010} quantum computing models allows one to consider the complexity associated specifically with quantum states without loss of generality.
We note that an arbitrary readout measurement can be always reduced to the computational basis measurement by adding the corresponding unitary to the circuit.
A similar trick can be applied to the consideration of measurement-based feed-forward operations, which can be replaced by controlled unitary gates and postponed measurements.
Thus, our approach based on the consideration of the complexity of quantum states is universal in the view of universality of the gate-based model of quantum computing.

\section{Classification of states} 

Let us start by reminding the fact that in order to achieve quantum computational advantage in general we work with quantum states, which are exponentially hard to simulate, i.e. predict  probabilities of measurement outcomes or at least mimic the corresponding sampling procedure.
If all the states we operate with are classically simulatable it seems unrealistic to expect quantum computational advantage in this case, so we stress on certain specific cases below.

Let us consider the following sets of $n$-qubit quantum states:
\begin{itemize}
    \item \textsf{Stab} is the set of stabilizer (``Clifford'') states, i.e. quantum states that are obtained with quantum circuits consisting only of Clifford gates applied to a computational basis state;
    \item \textsf{ClassSimMeas} is the set of states, for whose a classical algorithm of no more than polynomial in $n$ complexity that is able to reproduce the results of measurement (at least in a weak sense) of such states in the computational basis exists; 
    \item \textsf{ClassNonSimMeas} is the set of states, for which no classical algorithm of polynomial in $n$ complexity exists to reproduce the results of measurement (at least in a weak sense) of such states in the computational basis; 
    \item \textsf{QuantPrep}$_{1,2}$ is the set of states, which can be prepared on a quantum computer using no more than polynomial in $n$ in the length of the sequence of arbitrary single- and two-qubit gates applied to a computational basis state;
    \item \textsf{NotQuantPrep}$_{1,2}$ is the set of states that cannot be prepared on a quantum computer using a no more than polynomial in $n$ in size sequence of arbitrary single- and two-qubit gates applied to a computational basis state;
    \item \textsf{AreaLaw} (\textsf{VoumeLaw}) is the set of states, for whose the entanglement entropy of a region of space tends to scale, for large enough regions, as the size of the boundary (volume) of the region;
    \item \textsf{QuantCompAdv} is the set of states, which appear before the measurement in the computational basis for quantum algorithms with explicit circuits (e.g., without ``black box'' oracles) that have more than polynomial in $n$ advantage in comparison to the best (known or theoretically possible) classical algorithm.
\end{itemize}

We note that ${\sf ClassNonSimMeas}$ is a complement of ${\sf ClassSimMeas}$, as well as ${\sf QuantPrep}_{1,2}$ is a complement of ${\sf NotQuantPrep}_{1,2}$.
We would like to comment additionally with respect the \textsf{QuantPrep}$_{1,2}$ class. 
In its definition we stress on the possibility to express a general unitary transformation in terms of single- and two-qubit gates, which are available in practice. 
This class corresponds to the set of quantum states, which in practice can be prepared with a realistic quantum computer, whereas as it is known that typically decomposing $n$-qubit unitary $U$ into the sequence of single- and two-qubit gates generally requires sequence that is exponential in size.
This class can be extended to, let say, \textsf{QuantPrep}$_{1,2,\dots,{m}}$ in the case if up to $m$-qubit gates will be natively supported by a quantum hardware, which is yet not the case in practice. 
However, even if such gates exist for some fixed $m$, this does not influence on the asymptotic behavior of the length of the gates sequence.
The same holds for \textsf{NotQuantPrep}$_{1,2}$.

Recent studies from the side of quantum many-body and condensed matter physics, which are areas that heavily use the concept of simulation of quantum systems using classical approaches, have shown that the way, how the entanglement behaves with changing the bipartitioning of the whole system, plays a crucial role.
We note that this bipartitioning is usually realized with respect to the geometric topology of the corresponding physical system, e.g., one-dimensional chain or two- and three-dimensional array.
There are efficient classical tools to simulate entangled many-body quantum systems, where the entanglement of a region of space tends to scale, for large enough regions, as the size of the boundary of the region.
However, if such the entanglement grows as volume such methods cannot anymore work efficiently.
According to our definition, we refer these classes as \textsf{AreaLaw} and \textsf{VoumeLaw}, correspondingly. 
Specifically, tensor networks are good to approximate \textsf{AreaLaw} quantum states~\cite{Orus2019-3}, 
whereas neural-network quantum states (NNQS) are believed to be useful to describe certain \textsf{VoumeLaw} states~\cite{DasSarma2017,Carleo2022,Lvovsky2023}.
We also note ${\sf AreaLaw}$ and ${\sf VoumeLaw}$ are not complements of each other, since there can exist states whose entanglement scales neither as a volume nor as an area (e.g., be constant, or scale as some nontrivial power of the area).

\section{Relations between the sets of states} 

Here we would like to consider a number of statements and conjectures regarding the relations between the aforementioned classes of quantum states. 

By definition:

\begin{statement}
\textsf{QuantCompAdv} is in \textsf{QuantPrep}$_{1,2}$ and in \textsf{ClassNonSimMeas};
\end{statement} 

Only such states can give practical quantum advantage with realistic quantum computing devices.

\begin{statement}
In the view of the Gottesman-Knill theorem~\cite{Gottesman1998,NielsenChuang2000,Gottesman2004}, \textsf{Stab} is in \textsf{ClassSimMeas}.
\end{statement} 

This statement is obvious because of the possibility to simulate stabilizer quantum states with polynomial resources~\cite{Gottesman1998,NielsenChuang2000,Gottesman2004}. 

\begin{statement}
Due to results of~\cite{Gottesman2004,patel2003efficient,bravyi2021hadamard} \textsf{Stab} is inside \textsf{QuantPrep}$_{1,2}$.
Namely, the complexity of preparing a given stabilizer state $\ket{\Psi_n}$ scales as ${\cal O}(n^2/\log n)$.
\end{statement} 

This is due to the fact that stabilizer states can be efficiently prepared with single- and two-qubit gates, for example, 
via set of Pauli matrices and CNOT gates. 
An example of such state is the multiqubit GHZ state.

\begin{statement}
\textsf{Stab} crosses \textsf{AreaLaw} and \textsf{VoumeLaw}.
\end{statement} 

Examples of the states that exist on the crossing are between these classes are cluster state, which are used in measurement-based quantum computing~\cite{Briegel2001}. 

\begin{statement}
    \textsf{AreaLaw} crosses \textsf{ClassSimMeas}.
\end{statement} 

This statement comes from the fact tensor networks of the matrix-product-state (MPS) form can efficiently describe area-law entangled quantum states~\cite{Orus2019-3}.
MPS are particularly well suited for describing gapped 1D quantum lattice systems with local interactions~\cite{Hastings2006}.

\begin{conjecture}
    \textsf{AreaLaw} is in \textsf{ClassSimMeas}.
\end{conjecture} 

Here we conjecture that all area-law quantum states can be simulated with no more than polynomial classical resources.  

\begin{statement}
    \textsf{ClassSimMeas} crosses \textsf{VolLaw}.
\end{statement} 

This is due to the fact that certain \textsf{VolLaw} states can be efficiently described with NNQS~\cite{DasSarma2017,Carleo2022,Lvovsky2023}; 
for example, even for a 1D state with volume-law scaling MPS breakdowns, whereas NNQS works efficiently~\cite{DasSarma2017,Lvovsky2023}.

\begin{statement}
\textsf{QuantPrep} crosses \textsf{AreaLaw} and \textsf{VolLaw}.
\end{statement} 

This again can be demonstrated by the sequence of polynomial (${\sim}{n}^2$) sequence of two-qubit operations in the all-to-all connected qubit topology. 

\begin{conjecture}
There are exist \textsf{VolLaw} states that are out of \textsf{ClassSimMeas}.
\end{conjecture} 

Examples include noiseless quantum states that are prepared by random circuits, which can be used for demonstrating quantum computational advantage (it has been done for noisy circuits~\cite{Martinis2019,Pan2021-4}); 
in the noisy case, however, a polynomial-time classical algorithm for noisy random circuit sampling exists~\cite{Vazirani2023} (which, however, does not address aforementioned finite-size-circuit quantum advantage experiments).

\begin{conjecture}
\textsf{ClassSimMeas} crosses \textsf{notQuantPrep}$_{1,2}$.
\end{conjecture} 

\begin{figure}[t]
    \centering
    \includegraphics[width=0.99\linewidth]{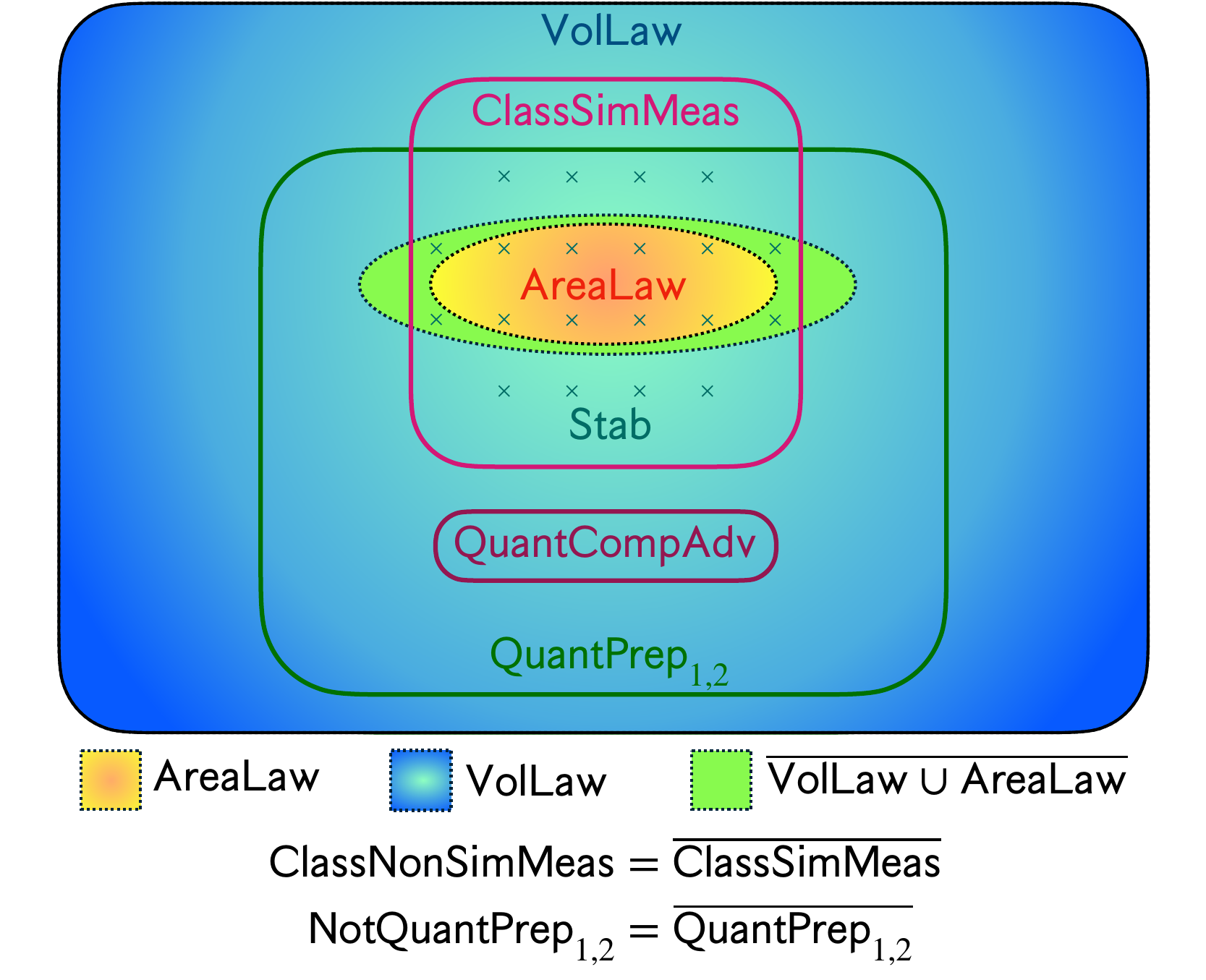}
    \caption{Representation of stated and conjectured relations between set of states.
    Overlines are used to denote complements. 
    We use a number of $\times$ symbols for  ${\sf Stab}$ to emphasize its finiteness for given $n$.
    }
    \label{fig:compl_classes}
\end{figure}

This conjecture is due to the fact that NNQS can simulate quantum states, which can be simulated classically, but at the same time cannot be prepared on a realistic quantum computer. 

The resulting relation between sets of states is sketched in Fig.~\ref{fig:compl_classes}.
Here we also would like to comment on the size of states sets for fixed value $n$.
It is known that the set of stabilizer states ${\sf Stab}$ contains 
\begin{equation}
    2^n\prod_{i=1}^n(2^i+1)
\end{equation} 
states~\cite{gross2006hudson}.
At the same time, ${\sf QuantPrep}_{1,2}$, ${\sf AreaLaw}$, ${\sf VolLaw}$, ${\sf NotQuantPrep}_{1,2}$ posses an infinite number of states, since adding a local continuous operation in the end of the state corresponding preparation circuit does not move states out of these sets.
${\sf ClassSimMeas}$ posses an infinite number of state due to intersection with ${\sf AreaLaw}$ and ${\sf QuantPrep}_{1,2}$.
An infinite subset inside ${\sf QunatCompAdv}$ contains state appeared in random circuits with, e.g., Haar-random single-qubit gates.
It is an interesting question whether it is possible to specify the class of problems within the definition of quantum computational advantage in such a way the corresponding set of states, or more precisely the set of corresponding read-out probability distributions, becomes finite.




\section{Outlook}

In this work, we have suggested an approach to establish relations between different classes of quantum states. 
We note that ``simplicity'' of states does not imply their impracticality for quantum information technologies in general, in particular, for quantum communication.
An illustrative example is the BB84 quantum key distribution protocol, where stabilizer states and Clifford measurements are used to solve a classically untractable problem of unconditionally secure key growing. 
Another intriguing area is oracle-based algorithms, which can provide a provable advantage (see, e.g., single-shot Bernstein-Vazirani algorithm~\cite{pokharel2023demonstration}). 
Yet, from our point of view, the potential speed-up may be related to quantum communication protocols, rather than to quantum computing, at least from the practical point of view.

\smallskip

\section*{Acknowledgments}

The authors thank N.N. Kolachevsky, B.I. Bantysh, and I.V. Ermakov for fruitful discussions. 
The research is supported by the Priority 2030 program at the National University of Science and Technology ``MISIS'' under the project K1-2022-027.

\bibliography{bibliography-rev.bib}

\end{document}